\documentclass[a4paper,twocolumn,longbibliography, accepted=2025-05-15]{quantumarticle}
\pdfoutput=1        
\usepackage[english]{babel}
\usepackage{graphicx}
\usepackage{xcolor}
\usepackage[colorlinks=true, citecolor=blue, urlcolor=blue ]{hyperref}
\usepackage{hyperref}
\usepackage{orcidlink}
\usepackage{amsmath,mathtools,verbatim}
\usepackage{amssymb}
\usepackage{dcolumn}  
\usepackage{bm} 
\usepackage{epstopdf}
\usepackage{booktabs}
\usepackage[caption=false]{subfig}
\usepackage{braket}
\usepackage{nicefrac}
\usepackage{soul}
\usepackage{float}
\usepackage[normalem]{ulem}
\allowdisplaybreaks

\begin{document}

\title{Protecting information in a parametrically driven hybrid quantum system}

\author{Siddharth Tiwary}
\email{siddharthtiwary@berkeley.edu}
\thanks{\orcidlink{0000-0002-3679-3749}}
\affiliation{Department of Physics, Indian Institute of Technology Bombay, Mumbai 400076, India}
\affiliation{Department of Physics, University of California, Berkeley, CA 94720, USA}

\author{Harsh Sharma}
\email{harsh.sharma@iitb.ac.in}
\thanks{\orcidlink{0000-0003-0136-2655}}
\affiliation{Department of Physics, Indian Institute of Technology Bombay, Mumbai 400076, India}

\author{Himadri Shekhar Dhar}
\email{himadri.dhar@iitb.ac.in}
\thanks{\orcidlink{0000-0002-5877-3415}\\ \\}
\affiliation{Department of Physics, Indian Institute of Technology Bombay, Mumbai 400076, India}
\affiliation{Centre of Excellence in Quantum Information, Computation, Science and Technology, Indian Institute of Technology Bombay, Mumbai 400076, India}

\thanks{The first two Authors contributed equally to this work.}

\begin{abstract}
The transfer and storage of quantum information in a hybrid quantum system, consisting of an ensemble of atoms or spins interacting with a cavity, is adversely affected by the inhomogeneity of the spins, which negates the coherent exchange of excitations between the physical components. Using a full quantum treatment based on variational renormalization group, we show how quantum information encoded in the states of a parametrically driven hybrid system is strongly protected against any decoherence that may arise due to the inhomogeneity in the spin-ensemble. 
\end{abstract}     
\maketitle

\section{Introduction\label{intro}}
The development of quantum devices must often accommodate conflicting requirements, ranging from readily interacting qubits for information processing and communication to stable, protected states for storage of information~\cite{Molmer2014}. An important cog in the contemporary quantum ecosystem that can overcome such conflicts is a hybrid system~\cite{Xiang2013,Kurizki2015}, where the synergy between different physical systems is achieved by delegating tasks to components that offer specific advantages. For instance, while superconducting circuits are amenable for information processing~\cite{Tsai1999}, an ensemble of atoms or spins can offer longer coherence times for storage~\cite{Julsgaard2004,Greze2014}. Such ensembles can then be coupled to resonators that readily allow for transfer of information~\cite{Verdu2009,Kubo2010,Schuster2010, Amsuss2011}. As such, hybrid systems based on spin-ensembles coupled to microwave cavities have gained significant traction in quantum information processing~\cite{Kubo2010,Schuster2010, Amsuss2011,Wesenberg2009,Clerk2020,Blais2021}. 

A key aspect of working with an ensemble of two-level atoms or spin-$1/2$ particles is that the transition frequencies for each particle may not be identical, a feature known as inhomogeneous broadening. Such broadening has the detrimental effect of decohering any information stored or transmitted from the ensemble to a coupled cavity, thus limiting the performance of any information processing protocol~\cite{Kurucz2011}. Over the years, several approaches have been proposed and implemented to overcome the effect of inhomogeneous broadening. A natural recourse is the ``cavity protection effect'', whereby the coupling between the spin-ensemble and the cavity is significantly increased to suppress the decoherence arising due to broadening~\cite{Diniz2011,Putz2014,Krimer2014,Sina2018,Sesin2023}. 
Spin-ensemble based hybrid systems already offer the advantage of operating in the strong-coupling regime, due to the enhanced collective coupling strength of a large number of spins~\cite{Kubo2010,Schuster2010, Amsuss2011}. However, increasing this coupling further to suppress decoherence may require operational regimes, such as high-Q cavities or increased spin density, which may either be difficult to implement or introduce unwanted spin interactions. 
Other approaches rely on refocusing mechanisms using spin-echo~\cite{Julsgaard2013} or sophisticated engineering of the spectral distribution of the ensemble based on either optimal selection~\cite{Bensky2012}, dynamical decoupling~\cite{Cai2012} or hole-burning~\cite{Krimer2015,Putz2017}. While each method has its advantages, they often rely on complex control protocols or challenging experimental schemes to manipulate the spins, which may impede the performance of the hybrid system.

In this work, we propose a radically different method to overcome decoherence due to inhomogeneous broadening -- one that does not involve altering the intrinsic spectral distribution of the spins but focuses on how information is encoded in the hybrid quantum system. In this approach, the quantum cavity is subjected to a parametric, two-photon drive, which exponentially enhances the coupling~\cite{You2015,Zeytinoglu2017} between the states of the cavity in the squeezed frame and the inhomogeneously broadened spin-ensemble.
Therefore, any information now encoded in this new cavity frame experiences an implicit cavity protection effect. A clear advantage here is that by tuning the parametric driving strength, the effective coupling can be increased without resorting to the design of expensive cavities or the use of a high spin-density ensemble. Such enhancement in coupling has been studied in other contexts, including ultrastrong coupling between a few qubits and the cavity~\cite{Qin2018,Leroux2018}, creation of entanglement~\cite{Wang2020,Chen2021a}, superradiant phase transition~\cite{Zhu2020,Yang2022} and squeezed lasing~\cite{Munoz2021}. Further, recent experiments have demonstrated such enhancement in different platforms, including trapped ions~\cite{Burd2021, Affolter2023, Burd2024} and superconducting circuits~\cite{Villiers2024}.

To analyze the decoherence and subsequent cavity protection effect during the temporal evolution of the hybrid quantum system, we first look at the semiclassical equations of motion for the average photonic and spin excitations~\cite{Bonifacio1982,Krimer2019,Zens2019}. However, the study of protection of quantum information encoded in the photonic states requires a full quantum treatment, which is done using a variational renormalization method that captures the temporal dynamics of a driven-dissipative, spin-ensemble-cavity based hybrid system~\cite{Dhar2018}. This then allows us to explicitly investigate any potential loss of information or fidelity of the encoded quantum state.

The paper is arranged in the following way. In Sec.~\ref{model}, we look at the model that describes the dynamics of the hybrid system consisting of an ensemble of spins inside a cavity under parametric driving. In Sec.~\ref{sc}, we derive the semiclassical equations for the photon and spin excitation operators and study the transition from weak to strong coupling regime. For full quantum solutions, we present the variational renormalization group method for driven-dissipative systems in Sec.~\ref{dmrg}, and show the protection effect under increasing driving strength. Finally, we end with a discussion of the main results in Sec.~\ref{disc}.

\section{The effective Hamiltonian\label{model}}
\begin{figure}[t]
\includegraphics[width=\columnwidth]{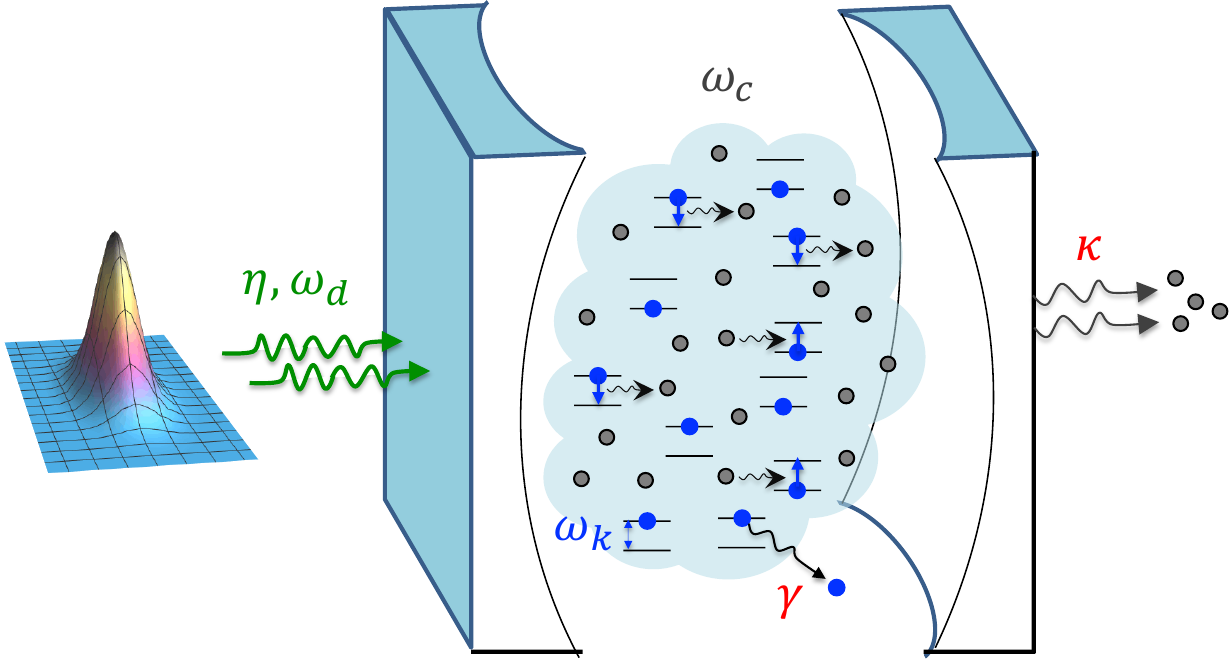}
\caption{A driven-dissipative hybrid quantum system, with $N$ spins in an ensemble interacting with a cavity, with a frequency $\omega_c$ and driven by a parametric, two-photon drive with frequency $\omega_d$ and intensity $\eta$. The $k^{th}$ spin in the ensemble has a transition frequency $\omega_k$,
and the spins and cavity lose excitations at rates $\gamma$ and $\kappa$, respectively.}
\label{schem}
\end{figure}

The dynamics of a hybrid quantum system, consisting of an ensemble of $N$ two-level atoms or spin-$1/2$ particles interacting with a quantum cavity, as shown in Fig.~(\ref{schem}), is governed by the Tavis-Cummings model~\cite{Tavis1968}. For a system subjected to a parametric, two-photon drive with a frequency $\omega_d$, the corresponding Hamiltonian ($\hbar =1$) under the rotating wave approximation is given by,
\begin{multline}
\mathcal{H}_0 = \frac{1}{2}\sum_{k=1}^N \Delta_k \hat{\sigma}_k^z + \Delta_c \hat{a}^\dag \hat{a} 
\\+  \sum_{k=1}^N g_k \left(\hat{\sigma}_k^- \hat{a}^\dag +\hat{\sigma}_k^+\hat{a}\right) -\frac{\eta}{2}(\hat{a}^2+\hat{a}^{\dag2}),
\label{tc}
\end{multline}
where the system is in a frame rotating with half the driving frequency, $\omega_d/2$. 
The cavity and the spin transition frequencies are given by $\omega_c$ and $\omega_k$ for $k=1,\dots,N$.
The corresponding frequency detunings with respect to $\omega_d/2$
are $\Delta_{c,k} = \omega_{c,k}-\omega_d/2$. 
As usual, $\hat{a}$ is the photon annihilation operator, $\{\hat{\sigma}_k^z,\hat{\sigma}_k^{\pm}\}$ are the spin operators given by the Pauli matrices, $g_k$ is the coupling between the $k^\textrm{th}$ spin and the cavity field, and $\eta$ is the strength of the parametric drive. 

For this study, the fundamental quantity of interest is the inhomogeneity of the spin-ensemble, given by the distribution of the spin transition frequencies $\omega_k$ and the spin-photon coupling $g_k$. The frequency distribution is taken to be
Gaussian~\cite{Marzin1994,Murray1995} (cf.~\cite{Sandner2012}) with a standard deviation or width given by $\delta$. For simplicity, the coupling is taken to be identical i.e., $g_k = g, \forall~ k$, which results in an effective coupling strength of $\Omega = \sqrt{\sum_{k=1}^N g_k^2} = \sqrt{N}g$. This demonstrates an enhancement of the coupling by a factor of $\sqrt{N}$ compared to that of a single spin. Hence, ensembles with a high number of atoms or spins readily exhibit phenomena related to strong light-matter coupling even when individual particles couple only weakly to the cavity~\cite{Kubo2010,Amsuss2011}.

Importantly, the effective spin-photon coupling can be further enhanced by parametrically driving the system. Using the unitary operator $\mathcal{U}=\exp\left[{r(\hat{a}^2-\hat{a}^{\dag2})/2}\right]$, where $r=1/2 \tanh^{-1}(\eta/\Delta_c)$ is the squeezing parameter, the cavity states can be transformed to a squeezed frame. The transformed spin-ensemble-cavity Hamiltonian is then given by $\mathcal{H}_{sq} = \mathcal{U}\mathcal{H}_0\mathcal{U}^\dag$, such that
\begin{multline}
    \mathcal{H}_{sq} = \tilde{\Delta}_c \hat{a}_s^\dag \hat{a}_s + \frac{1}{2}\sum_{k=1}^N\left\{ \Delta_k \hat{\sigma}_k^z \right. \\
+  g_k e^r \left(\hat{a}_s+\hat{a}_s^\dag\right)\left(\hat{\sigma}_k^- +\hat{\sigma}_k^+\right)  \\
- \left. g_k e^{-r} \left(\hat{a}_s-\hat{a}_s^\dag\right) \left(\hat{\sigma}_k^- -\hat{\sigma}_k^+\right)\right\},
\label{dicke}
\end{multline}
where {$\tilde{\Delta}_c = \Delta_c /\cosh{(2r)}$} {and $\hat{a}_s = \cosh{(r)} \hat{a} + \sinh{(r)} \hat{a}^\dag$ is the cavity operator in the squeezed frame}. 
{For large $r$, such that $e^{-r}\to 0$, the Hamiltonian $\mathcal{H}_{sq}$ is nothing but the quantum Dicke model or the $N$-spin Rabi model with inhomogeneous spins and an exponentially enhanced coupling $\tilde{g}_k = g_k e^{r}/2$. The system is in resonance if the mean spin detuning $\bar\Delta_k = \sum_k \Delta_k/N \approx \tilde{\Delta}_c$. For $\tilde{g}_k \approx \tilde{\Delta}_c$, the system is in the ultra-strong coupling regime~\cite{Ciuti2005,Kockum2019}.
However, for $\tilde{g}_k \ll \tilde{\Delta}_c$, the rotating wave approximation is valid and $\mathcal{H}_{sq}$ reduces to the Tavis-Cummings model with collective coupling $\tilde{\Omega} = \Omega \cosh{(r)}$. As such, the 
system can be studied in different operational regimes by manipulating the 
effective spin-cavity coupling, which can be readily controlled by changing the parametric driving amplitude $\eta$.}

\section{Dynamics in the semiclassical regime\label{sc}}
A key indicator to investigate the protection of information encoded in hybrid quantum systems is to investigate how quickly the average spin excitation or photon number in the cavity is lost as the system evolves. These quantities can be readily calculated using semiclassical equations of motion~\cite{Bonifacio1982,Krimer2019,Zens2019}.  
Now, in addition to inhomogeneity in the ensemble, the spins and the cavity in a hybrid system are also intrinsically lossy, i.e., they naturally lose coherence with time. Thus, the dynamics of a hybrid system is best described by a Lindblad master equation (ME), given by 
\begin{multline}
    \frac{d\rho}{dt} = -i[\mathcal{H},\rho] + \kappa \mathcal{L}_{\hat{a}}[\rho] \\
    + \sum_{k=1}^N \{\gamma_h \mathcal{L}_{\hat{\sigma}^-_k}[\rho]
    + \gamma_p \mathcal{L}_{\hat{\sigma}^z_k}[\rho]\},
\label{lind}
\end{multline}
where $\rho$ is the density matrix of the system and the Lindblad operators are $\mathcal{L}_{\hat{x}}[\rho]=\hat{x}\rho \hat{x}^\dag-\frac{1}{2}\{\hat{x}^\dag \hat{x},\rho\}$. The photon loss rate is $\kappa$, the rate of radiative decay and dephasing of spins is $\gamma_h$ and $\gamma_p$. 
For the Hamiltonian in Eq.~(\ref{tc})$, \mathcal{H}=\mathcal{H}_0$, the
equations of motion for the photonic and spin excitations, $\langle \hat{a}\rangle$, $\langle \hat{\sigma}_k^-\rangle$ and $\langle \hat{\sigma}_k^z\rangle$ are calculated using Eq.~(\ref{lind}). For large $N$, under the semiclassical or mean-field approximation, all correlations between the spins and the cavity can be ignored~\cite{Bonifacio1982,Krimer2019,Zens2019}. This allows factorization of all first-order correlations, i.e., $\langle \hat{\sigma}_k^-\hat{a}\rangle \approx \langle \hat{\sigma}_k^- \rangle \langle \hat{a}\rangle$, in the equations of motion of the hybrid system and we get a closed set of equations,
\begin{subequations}\label{semiclass}
    \begin{align}
        \frac{d\braket{\hat{a}}}{dt} &= -(\kappa+i{\Delta}_c)\braket{\hat{a}}-i\sum_kg_k\braket{\hat{\sigma}_k^-}+i\eta\braket{\hat{a}}^*,\nonumber\\
        \frac{d\braket{\hat{\sigma}_k^-}}{dt} &=-(\gamma_h+2\gamma_p+i\Delta_k)\braket{\hat{\sigma}^-_k}+ig_k\braket{\hat{\sigma}^z_k}\braket{\hat{a}},\nonumber
    \end{align}
     \vspace{-0.3cm}
    \begin{multline}
        \frac{d\braket{\hat{\sigma}_k^z}}{dt} = -2\gamma_{h}(1+\braket{\hat{\sigma}^z_k}) + 2ig_k(\braket{\hat{\sigma}_k^-}\braket{\hat{a}}^*-\braket{\hat{\sigma}_k^+}\braket{\hat{a}}),\\~\tag{4}
    \end{multline}
\end{subequations}
where {$\braket{a^\dag}=\braket{a}^*$ and $\braket{\hat{\sigma}_k^+}=\braket{\hat{\sigma}_k^-}^*$}. By solving the equations above, we study the dynamics of a hybrid system consisting of an inhomogeneous ensemble of $N = 10^4$ spins. This ensemble is sufficiently large to use the semiclassical approximation in the stable region~\cite{Zens2019}, {and to drive the system in linear regime, we ensure that $\tilde{g}_k \ll \tilde{\Delta}_c$}. Moreover, to {characterize only} the effect of inhomogeneous broadening in the semiclassical limit, the photon and spin losses are taken to be {very small}, as compared to the {spin-cavity coupling} and width of the distribution. 

\begin{figure}[t]
\centering
\includegraphics[width=3.2in]{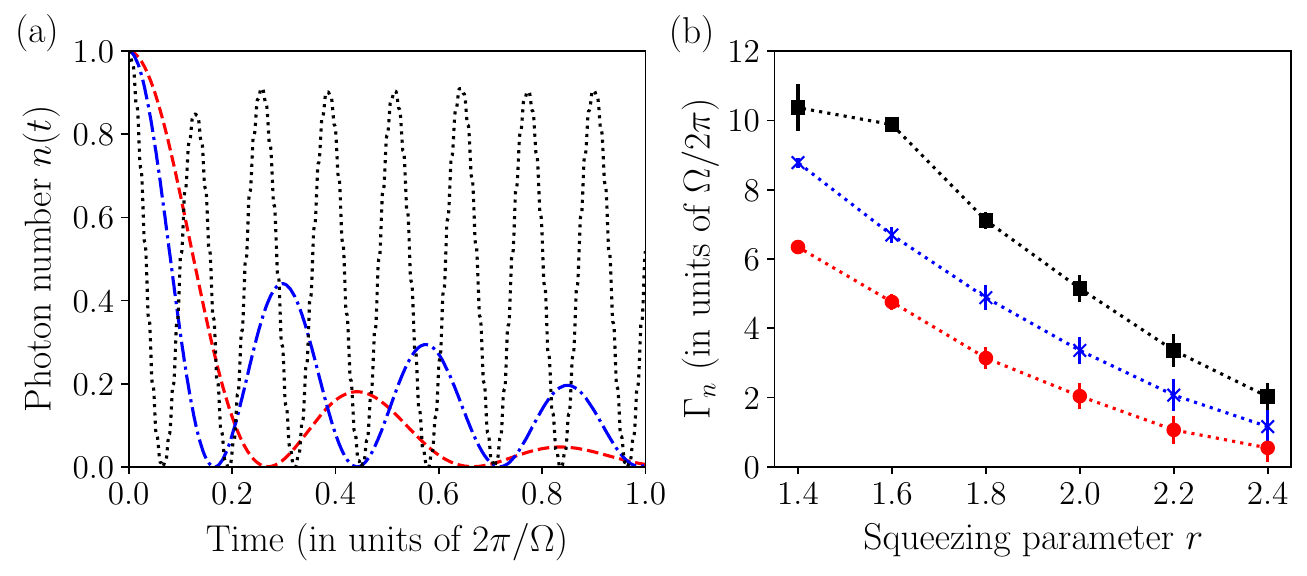}
\caption{{Semiclassical dynamics of an ensemble of spins inside a cavity. The figures show (a) the time evolution of the average photon number $n(t)=|\braket{\hat{a}_s}|^2$. The effective coupling is $\Omega \approx 6 \times 10^{-4}~\Delta_c$, and the spin detuning has a Gaussian frequency distribution of width {$\delta = 0.75~\Omega$} and mean $\Bar{\Delta}_k = \tilde{\Delta}_c$, and squeezing parameter $r=0.0$ (red dashed), $r=1.0$ (blue dot-dashed), and $r=2.0$ (black dotted), and (b) the variation in decay rate $\Gamma_{n}$ with $r$, for width $\delta = 0.75~\Omega$ (red circle), $\delta = 1.75~\Omega$  (blue cross), and $\delta = 2~\Omega$ (black square). The error bars show the standard deviation between the actual value of $n(t)$ and those corresponding to the fitted decay rate $\Gamma_{n}$. All axes are dimensionless.}
}
\label{fig:semi}
\end{figure}

Figure~\ref{fig:semi} captures the dynamics of a hybrid system under parametric driving. The semiclassical equations are derived using the Hamiltonian $\mathcal{H}_0$, such that the average photon number in the squeezed frame is given by $n(t) = |\braket{\hat{a}_s}|^2=\cosh(2r)|\braket{\hat{a}}|^2 - \sinh(2r)\Re(\hat{a}^2)$, where $r$ is the squeezing parameter. 
Similar results are obtained if one instead studies the equations of motion in the {squeezed frame, by using $\mathcal{H}=\mathcal{H}_{sq}$ in Eq.~(\ref{lind}).} In Fig.~\ref{fig:semi}(a), the temporal evolution of $n(t)$ is shown for different values of squeezing parameter $r$, and {for} a fixed width of the spin {detuning distribution $\delta = 0.75~\Omega$.} It is assumed that the hybrid system is initialized such that the cavity has photon number equal to unity i.e., $n(0) =1$. 

The figure shows that for $r=0$, the average photon number quickly decoheres with time due to the inhomogeneity in the spin-ensemble (other sources of dissipation {are negligible} at this point). However, these excitations are sustained for higher values of $r$, which shows that boosting the parametric drive leads to a reduced decay rate and a longer lifetime for the photon.
The rate $\Gamma_{n}$ at which the cavity excitation $n(t)$ is decohering can be captured by fitting the peaks of the Rabi oscillations using the relation, $n(t) = n(0) e^{-\Gamma_{n} t}$. So, higher $\Gamma_{n}$ implies faster decoherence of the photonic excitations in the hybrid quantum system. Figure~\ref{fig:semi}(b) shows the variation of the rate of decay $\Gamma_{n}$ as a function of squeezing parameter $r$, for spin-ensembles of different width $\delta$. The plots clearly show that loss of excitation and decoherence in the system is more severe for ensembles with more inhomogeneity (larger width $\delta$). Importantly, parametric drives of increasing intensities, which give us higher values of $r$, are able to reduce the decay rate $\Gamma_{n}$ successfully even for significantly broadened spin-ensembles, thus giving an overall ``cavity protection'' effect. In fact, the protection is also observed when the inhomogeneity is larger than the effective coupling, i.e., $\delta > \Omega$, albeit in the absence of any other dissipation in the system.

To better understand the cavity protection effect, the net loss or decay in the system needs to be estimated. This includes losses due to inhomogeneous broadening and losses in the system such as photon decay $\kappa$, spin decay $\gamma_h$ and dephasing $\gamma_p$.  In the semiclassical regime, for large $N$ and $\kappa,\gamma_h,\gamma_p \ll \tilde{\Omega}$, the average photon number in the cavity decays with the rate~\cite{Diniz2011}, 
\begin{equation}
     \Gamma_{n}(\tilde{\Omega}) = \kappa + \gamma_h + 2 \gamma_p + \pi\rho(\tilde{\Omega})\tilde{\Omega}^2,
\end{equation}
where $\rho(\tilde\Omega)\sim 2^{-\tilde\Omega^2/\delta^2}/\delta$, is the Gaussian spectral density of atoms or spins in the ensemble, with width $\delta$. The term  $\rho(\tilde\Omega)\tilde{\Omega}^2$ governs the decoherence induced by inhomogeneous broadening, where $\tilde{\Omega} = \Omega \cosh{(r)}$ is the effective coupling strength in the squeezed frame. 
For a fixed width $\delta$, the cavity protection effect is directly related to $\tilde{\Omega}$, which is a function of the squeezing parameter $r$. For $\tilde{\Omega} \gg \delta$, there is a protective energy gap between the superradiant and the subradiant spin-wave modes, which does not allow inhomogeneity to induce decoherence in the system~\cite{Kurucz2011}. 
As a result, by increasing $r$, the net decay rate $\Gamma_n$ can be suppressed and eventually the dynamics of the system becomes independent of inhomogeneous broadening. To quantify the suppression in errors due to inhomogeneity, we define the term
\begin{equation}
    \xi_r \equiv \frac{\Gamma_n(\tilde{\Omega})}{\Gamma_n(\infty)} - 1 = \frac{\pi\rho(\tilde\Omega)\tilde\Omega^2}{\kappa + \gamma_h + 2 \gamma_p}.
    \label{CavEnhancement}
\end{equation}
Note that $\xi_r$ becomes negligible in the limit $r\to \infty$ and will therefore serve as a good figure of merit in understanding cavity protection effect in inhomogeneous spin-ensembles.

\section{Protecting information in the hybrid system\label{dmrg}}
While semiclassical equations highlight the revival of macroscopic quantities such as the average photon number in the cavity, to study protection of {properties such as correlations or quantum information encoded in photonic states}, the dynamics of the full quantum system is necessary. This is a computationally challenging task due to the large Hilbert space of the spin-ensemble-cavity system. To overcome this, we use a variational renormalization group method to study driven-dissipative dynamics~\cite{Dhar2018}, {which is a tensor-network method similar to density matrix renormalization group and matrix-product states~\cite{Schollwoeck2011}. 
The method employs a time-adaptive approach} to temporally evolve the density matrix $\rho$ of a spin-cavity system of about hundred spins, using the Lindblad ME formalism. To achieve this, all states $\rho$ and operators $\hat{O}$ are mapped to a higher dimensional space as superkets $|\rho\rangle\rangle$ and superoperators $\hat{\mathcal{O}}$ i.e., $\rho \rightarrow \textbf{vec}(\rho)=|\rho\rangle\rangle$ and $\hat{O}\rho \rightarrow (\hat{O}\otimes \mathbb{I})|\rho\rangle\rangle = \hat{\mathcal{O}}|\rho\rangle\rangle$. This allows the master equation in Eq.~(\ref{lind}) to be mapped to a Schr{\"o}dinger like equation, $d|\rho\rangle\rangle/dt = \tilde{\mathcal{L}}|\rho\rangle\rangle$, in the superoperator space, where 
\begin{multline}
\tilde{\mathcal{L}} = -i(\mathcal{H} \otimes \textbf{I}-\textbf{I} \otimes \mathcal{H}^T) + \kappa \tilde{\mathcal{L}}_{\hat{a}} \\
+ \sum_k\{\gamma_h \tilde{\mathcal{L}}_{\hat{\sigma}_k^-} + \gamma_p \tilde{\mathcal{L}}_{\hat{\sigma}_k^z}\},
\label{sup_lind}
\end{multline}
and $\tilde{\mathcal{L}}_{\hat{x}} = \hat{x}\otimes\hat{x}^*-\frac{1}{2}\hat{x}^\dag\hat{x}\otimes\textbf{I}-\frac{1}{2}\textbf{I}\otimes\hat{x}^T\hat{x}$. At this point, the variational renormalization has two key parts. First, finding the renormalized representation for the superket $|\rho\rangle\rangle$ in a significantly reduced subspace, and second, evolving the initial state $|\rho\rangle\rangle$ in a time-adaptive manner. As with most tensor-network methods, the renormalized space is obtained by eliminating the null space or those with marginal singular values. 
The time evolution is then governed using the Lindblad superoperator in Eq.~(\ref{sup_lind}), using the transformed Hamiltonian $\mathcal{H}=\mathcal{H}_{sq}$, and evolved adaptively for small intervals $\Delta t$, such that $\ket{\rho(t+\Delta t)}\rangle = e^{\tilde{\mathcal{L}}\Delta t}\ket{\rho(t)}\rangle$. Now, $\tilde{\mathcal{L}} = \sum_k \tilde{\mathcal{L}}_k$, and the
second-order Suzuki-Trotter decomposition helps expand $e^{\tilde{\mathcal{L}}\Delta t}$ into a product of single spin-cavity evolution: $e^{\tilde{\mathcal{L}}_N \Delta t/2}e^{\tilde{\mathcal{L}}_{N-1}\Delta t/2}\dots e^{\tilde{\mathcal{L}}_1 \Delta t}\dots e^{\tilde{\mathcal{L}}_{N-1} \Delta t/2}e^{\tilde{\mathcal{L}}_{N}\Delta t/2}$ (see Refs.~\cite{Dhar2018,Zens2021}). The computation then follows an iterative sweeping protocol as used in other time-adaptive tensor-network methods~\cite{Daley2004,White2004}. 

{An important impact of parametrically driving an open cavity is that its environment is also affected. In other words, squeezing the cavity also transforms the Lindblad operator $\mathcal{L}_{\hat{a}}$, giving rise to additional noise terms~\cite{Qin2018} in the master equation. For instance,
\begin{multline}
    \mathcal{L}_{\hat{a}}[\rho] \rightarrow  (n_{\text{th}}+1)\mathcal{L}_{\hat{a}_s}[\rho] + n_{\text{th}} \mathcal{L}_{\hat{a}_s^\dag}[\rho]  \\ 
    - m_{\text{2-ph}} \mathcal{L}'_{\hat{a}_s}[\rho] - m_{\text{2-ph}}^* \mathcal{L}'_{\hat{a}_s^\dag}[\rho],
\end{multline}
where $\mathcal{L}'_{\hat{x}}[\rho]=\hat{x}\rho \hat{x}-\frac{1}{2}\{\hat{x}^2,\rho\}$. Here, $n_{\text{th}}$ and $m_{\text{2-ph}}$ are functions of squeezing parameter $r$, and correspond to thermal noise and two-photon correlations introduced by squeezing, respectively. 
However, these additional errors can be eliminated by coupling the cavity mode to a squeezed-vacuum reservoir with the same squeezing parameter $r$, but phase-shifted by $\pi$ relative to the two-photon drive.}

\begin{figure}[t]
    \centering
    \includegraphics[width=0.425\textwidth]{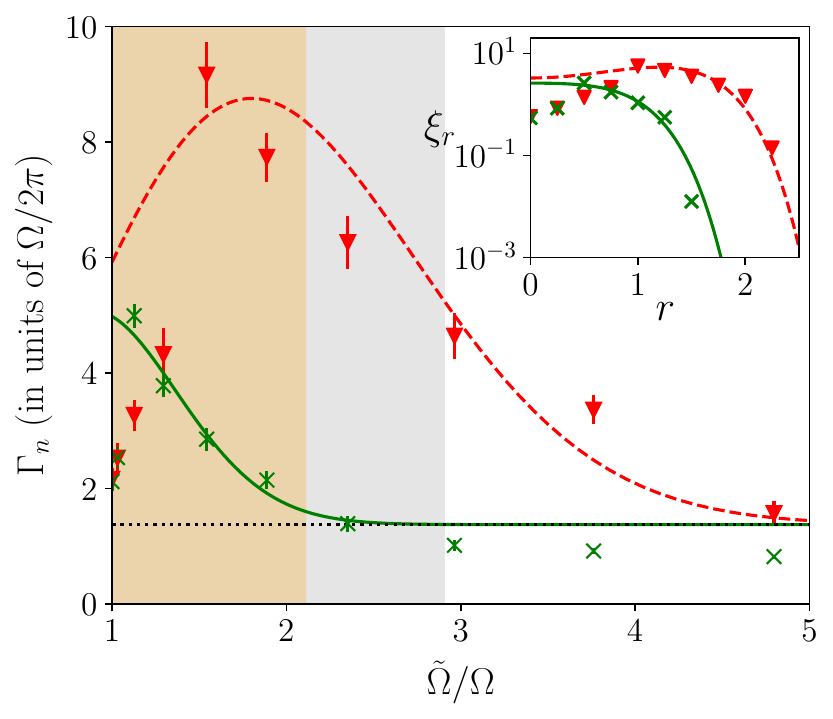}
    \caption{{Cavity protection effect in a hybrid system with inhomogeneous spins. The plots show the variation of decay rates $\Gamma_{n}$, for an inhomogeneous ensemble with width $\delta = 0.75~\Omega$ (green crosses) and $\delta = 1.5~\Omega$ (red triangles), obtained from numerical calculations for $N=100$ spins. The solid-green and dashed-red curves correspond to the theoretical value of $\Gamma_{n} (\tilde{\Omega})$ in the semiclassical limit, with the dotted line marking the theoretical minimum decay rate $\Gamma_{n}(\infty)$. The shaded region corresponds to squeezing regimes $r \leq 1.38$~\cite{Burd2024} (orange) and $r\leq 1.73$~\cite{Vahlbruch2016} (gray) that have been achieved experimentally. The inset shows the variation of $\xi_r$, where the markers are same as above. The error bars are the same as in Fig.~\ref{fig:semi}. Both axes are dimensionless.}}
    \label{Fig:decRates}
\end{figure}

For the full quantum dynamics of the hybrid system using variational renormalization group, {an ensemble of $N = 100$ spins is considered inside the cavity. For consistency with semiclassical results, the same set of parameters are used}, including the coupling strength $\Omega$. However, we now consider other dissipative terms, such as the photon loss rate $\kappa \approx 10^{-4}~\Delta_c$, and radiative loss and dephasing for spins equal to $\gamma_h = \kappa/8$ and $\gamma_p = \kappa/16$, respectively.
The loss rates we consider here are comparable to those observed across different experimental platforms such as transmon qubits~\cite{Minev2016,Villiers2024}, trapped ions~\cite{Burd2021,Affolter2023,Burd2024} and NV centres~\cite{Atac2009,Kubo2010,Putz2014,Putz2017}. A detailed comparison of the key parameters is shown in Table~\ref{Tab:Params} in 
Appendix~\ref{App1}.

{Figure \ref{Fig:decRates} demonstrates cavity protection effect by highlighting the reduction of the decay rate of average photon number $\Gamma_n$ as a function of $\tilde{\Omega}$, the effective coupling strength in the squeezed frame.
For $\tilde{\Omega} > \delta$, the numerically obtained values of $\Gamma_n$ from the full quantum dynamics of $N=100$ spins are close to the theoretical predictions from the semiclassical limit, for both $\delta = 0.75~\Omega$ and $\delta = 1.5~\Omega$. 
The results show a 2.5 fold reduction in $\Gamma_n$ for width $\delta \approx 0.75~\Omega$, for squeezing
parameter $r = 1.5$, and by a factor of 6 for $\delta \approx 1.5~\Omega$ and $r = 2.25$.
The inset of Fig.~\ref{Fig:decRates} exhibits the suppression of error by reducing the factor $\xi_r$ by a factor of $10^3$ and reducing the decay rate close to $\Gamma_n(\infty)$ at $r=1.75$ for  $\delta \approx 0.75~\Omega$. Similar reduction is also observed for higher $\delta$, albeit at higher squeezing $r$, which is consistent with the protection effect~\cite{Diniz2011}.
The ``cavity protection'' arising from parametric driving deviates from the semiclassical limit in some aspects. Firstly, for small $r$, the dynamics of the system is affected by the term $g_ke^{-r}$ in Eq.~(\ref{dicke}). Moreover, for $\tilde{\Omega}<\delta$, the system can become weakly coupled, which results in increase in $\Gamma_n$ for small $\tilde{\Omega}$~\cite{Putz2014}, as observed for $\delta = 1.5~\Omega$ in Fig.~\ref{Fig:decRates}. 
Secondly, for high $r$ and $\tilde{\Omega}$, the decay rate $\Gamma_n$ can reduce below the semiclassical lower bound $\Gamma_n(\infty)$, as observed for $\delta = 0.75~\Omega$. This is due to the finite size of the ensemble (or small $N$), which can lead to the creation of collective dark states as observed in experiments using spectral hole burning~\cite{Putz2017}.}

{The full quantum dynamics of the hybrid system allows for the study of protection of quantum information via parametric driving.}
At the start, the quantum information is encoded in the effective state of the cavity in the squeezed frame, which at time $t=0$ is given by $\ket{\psi_c(0)}$. As the dynamics is governed by the Lindblad ME, the cavity state at {any} time $t$ is given by the density matrix $\rho_c(t)$. 
To study the effect of decoherence and protection {of the encoded quantum information}, we look at two distinct figures of merit, viz. quantum fidelity and the Wigner function. The quantum fidelity $\mathcal{F}(t)$ is taken between the state $\rho_c(t)$ at $t$ and the initial state encoded in the system $\ket{\psi_c(0)}$, i.e., {$\mathcal{F}(t) = \sqrt{\langle \psi_c(0)|\rho_c(t)|\psi_c(0)\rangle}$}.
The Wigner function allows us to study the phase-space properties of the encoded information, which is not captured by the fidelity, and is defined as $W(p,q)=\frac{1}{\pi}\int e^{2iqy} dy \langle p-y|\rho_c(t)\ket{p+y}$.
{As it maps the density matrix in Hilbert space to a distribution in phase space, the function contains the full information present in the quantum state. As such, any substantial loss of information due to decoherence can be readily visualized by the change in its Wigner function in the phase space.}
\begin{figure}[t]
\centering
\includegraphics[width=3.3in]{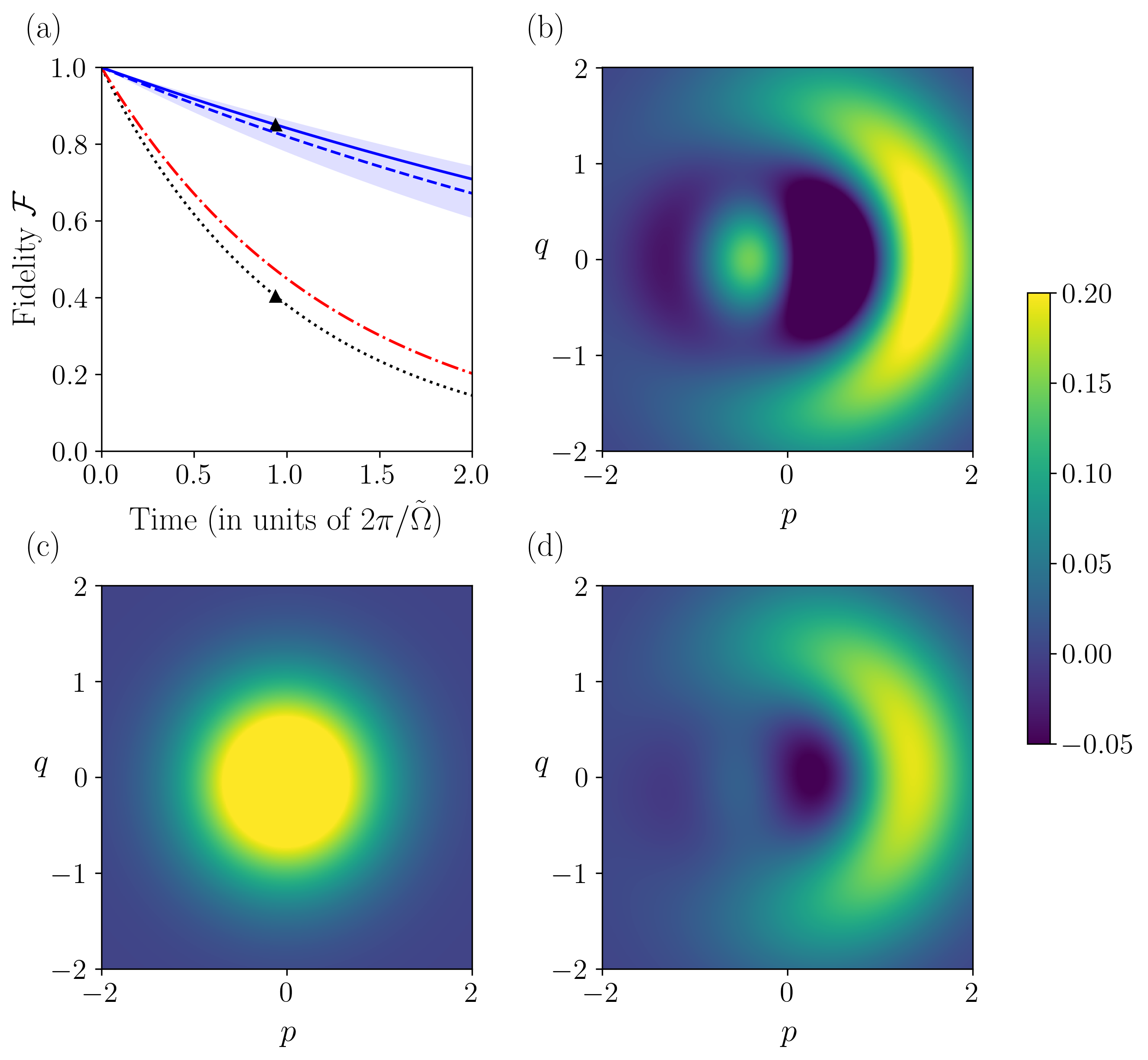}
\caption{Dynamics of quantum information encoded in a hybrid system consisting of $N=100$ spins inside a cavity. The width of the frequency distribution is $\delta = 0.75~\Omega$.
The plots in (a) show the evolution of the fidelity $\mathcal{F}$ between the effective cavity state at time $t$ and the initial state $|\psi_c(0)\rangle$ for squeezing parameter $r=0.0$ (dotted black), $r=1.0$ (dot-dashed red), and $r=1.75$ (solid blue). {The average fidelity (dashed blue) calculated over 540 Haar random initial states is also shown, with the shaded area representing unit standard deviation.} 
The other subfigures show the Wigner function $W(p,q)$ of the initial cavity state $|\psi_c(0)\rangle$ in (b), and the state at $t$ corresponding to fidelity (black triangles in (a)) for parameters (c) $r=0.0$ and (d) $r=1.75$.}
\label{fig_dmrg}
\end{figure}

Figure~\ref{fig_dmrg} shows the fidelity of the information encoded in the initial state and the temporally evolving photon density matrix $\rho_c(t)$, as well as its quasi-probability distribution in terms of $W(p,q)$. 
{The cavity at $t=0$ encodes the superposition state }$\ket{\psi_c(0)}=\frac{1}{\sqrt{2}}\{\ket{1}+\ket{2}\}$, where $\ket{1}$ and $\ket{2}$ are the energy eigenstates in the transformed basis. {While loss in excitations given by $\kappa$ and $\gamma_h$ are more (less) effective on states with higher (lower) excitation, dephasing ($\gamma_p$) acts strongly on states with higher coherence. As such we consider the maximally coherent state  with photon number equal to the mean excitation.}
In Fig.~\ref{fig_dmrg}(a), the plot shows the best-fit envelope of the fidelity of the density matrix at $t$ with the initially prepared superposition state that encodes the information~\cite{Ecomment}. 
The spin-ensemble has an inhomogeneity given by the width of the distribution $\delta = 0.75~\Omega$.
As such, the decrease in $\mathcal{F}(t)$  corresponds to decoherence or loss of information, as the initially encoded state is no longer accurately accessible. 
However, under increased parametric driving ($r=1.0$ and $1.75$), the loss of fidelity is significantly minimized compared to the undriven system ($r=0$), thus highlighting a robust protection effect on the information encoded in the system. 
This is also evident in the phase-space properties of the information stored in the cavity as shown in Figs.~\ref{fig_dmrg}(b)-(d), in terms of the Wigner function of $\rho_c(t)$. While Fig.~\ref{fig_dmrg}(b), shows the $W(p,q)$ for the initially encoded state, Figs.~\ref{fig_dmrg}(c) and (d) correspond to density matrices after one Rabi cycle for the undriven ($r=0$) and the parametrically driven ($r=1.75$) system, respectively. These plots clearly show that the phase coherence is better preserved for higher values of $r$.

{Figure~\ref{fig_dmrg}(a), also shows the evolution of the average fidelity, calculated over 540 Haar random initial states, and its temporal behaviour is quite close to that of the maximally coherent initial state.} 
Notably, for $r=0$, the fidelity drops below ${1}/{2}$ after one Rabi oscillation ($t\sim 1$), which is the probability of randomly guessing the stored information. However, under parametric driving, {at large $r$ the fidelity dies down at a rate governed by the losses in the system, while inhomogeneous broadening is suppressed.} This highlights the strong protection experienced by the information stored in the hybrid quantum system. 

\begin{table*}[t]
\centering
\begin{tabular}{p{0.185\linewidth}|p{0.315\linewidth}|ccccc}
\toprule
Reference   & Remarks   & $\kappa/\omega_c$ & $\gamma_h/\omega_c$  & $\gamma_p/\omega_c$  & $\delta/\omega_c$ & $\Omega/\omega_c$ \\ \hline 
~    & ~   & ~ & ~ & ~ & ~ & ~ \\
Our simulations     & --    & $10^{-4}$ & $\approx 10^{-5}$ & $\approx 10^{-6}$ & $\approx 10^{-3}$ & $\approx 10^{-3}$ \\
~    & ~   & ~ & ~ & ~ & ~ & ~ \\
Minev et. al.~\cite{Minev2016}  & Transmon qubit coupled to SC resonator with two modes & $\approx 10^{-4}$ & $\approx 10^{-5}$  & $\approx 10^{-5}$  & --  & $\approx 10^{-3}$  \\
A. Imamo\ifmmode \breve{g}\else \u{g}\fi{}lu~\cite{Atac2009} & spin-ensemble coupled to SC cavity with a buit-in transmon qubit which  provides non-linearity  & $\approx 10^{-4}$ & $\leq 10^{-4}$        & --                   & --   & $\approx 10^{-3}$ \\ 
Kubo et.al.~\cite{Kubo2010} & NV centre ensemble coupled to SC resonator & $\approx 10^{-4}$ & $\approx 10^{-3}$      & --                   & --                & $\approx 10^{-2}$ \\ 
Putz et. al.~\cite{Putz2014, Putz2017}  & NV centre ensemble coupled to SC cavity & $\approx 10^{-4}$ & $\approx 0$ & --    & $\approx 10^{-2}$ & $\approx 10^{-2}$ \\ 
Villiers et. al.~\cite{Villiers2024}    & Transmon qubit coupled to a waveguide oscillator & $\approx 10^{-3}$ & $\approx 10^{-3}$ & $\approx 10^{-3}$ & -- & $\approx 10^{-3}$ \\
Burd et. al.~\cite{Burd2021}    & Two Mg$^{2+}$ ions trapped using radio frequency trap & $\approx 10^{-6}$ & $\approx 10^{-4}$ & -- & -- & $\approx 10^{-3}$ \\
Diniz et. al.~\cite{Diniz2011}  & Cavity protection effect theory & $10^{-4}$ & $10^{-7}$ & -- & $\approx 10^{-3}$ & $\approx 10^{-2}$ \\
\bottomrule
\end{tabular}%
\caption{Comparison of parameters taken in various studies. The experimental demonstration of cavity protection effect is shown in Refs.~\cite{Putz2014,Putz2017}, while the setups in Refs.~\cite{Burd2021,Villiers2024} implement parametric driving. References~\cite{Minev2016,Atac2009,Kubo2010} are based on circuit-QED experiments.}
\label{Tab:Params}
\end{table*}

\section{Discussion\label{disc}}
Hybrid quantum systems based on spin-ensembles interacting with a cavity are ubiquitous in the design of quantum computing and communication architecture. As such, quantum information stored or transferred using these systems needs to be protected from decoherence arising due to inhomogeneity in the system.
While there exist several sophisticated techniques to regain coherence, based on engineering of the spin distribution or using expensive cavities, the present results show that by simply using a parametric drive on the hybrid system and encoding information in the transformed states, an effective protection protocol can be achieved. 

{A pertinent example is the experimental demonstration of cavity protection effect in an NV centre spin-ensemble coupled to a superconducting resonator~\cite{Putz2014}, where a 1.4 times increase $\Omega$ is experimentally achieved. 
Moreover, it was estimated that a 3-fold increase could be achieved by packing the mode volume with NV centres, but only at the cost of introducing finite spin-spin interaction.
In our scheme, the additional enhancement can be achieved using a parametric drive 
with squeezing factor $r \approx 0.81$, which has been experimentally realized in superconducting circuits~\cite{Villiers2024}.  Importantly, this allows strong cavity protection effect without giving rise to spin-spin interactions that may adversely affect the coherent exchange of excitations.
Similar parametric driving, with squeezing parameter $r \approx 1$, has also been demonstrated in experiments with trapped ions~\cite{Affolter2023}. Further enhancement can be achieved by increased squeezing up to $r \approx 1.8$, which is close to what can be achieved in state-of-the-art optical experiments~\cite{Vahlbruch2016}. In superconducting quantum interference devices (SQUID), squeezing of $r \approx 2$ has been reported~\cite{Yamamoto2008}.}

An important aspect of the method is that it is applicable across different frequency regimes and platforms. For example, parametric driving in platforms such as trapped ions can experimentally achieve squeezing as high as $r \approx 1.4$~\cite{Burd2024}, using a pumping amplitude $\eta$ in the order of 1~MHz. In experiments with transmon qubit~\cite{Villiers2024}, squeezing of $r\approx 0.81$  can be experimentally achieved for $\Delta_c/2\pi = 20$~MHz and $\eta/2\pi = 17$~MHz. While parametric amplification in higher frequency regimes are difficult to achieve in nonlinear materials~\cite{Yanagimoto2022}, it is more amenable in superconducting circuits by controlling flux passing through a SQUID~\cite{Yamamoto2008} or, more recently, using
highly anharmonic flux qubits called quartons~\cite{Ye2021}.

It is to be noted that our protocol is significantly different from other applications of squeezed light, such as its use for enhanced precision measurements~\cite{Xiao1987,Schnabel2017} or encoding information in continuous variable information processing~\cite{Braunstein2005,Takeda2013,Park2022}. From an experimental perspective, the use of higher-order interactions in hybrid systems can be naturally adapted in the design of qubits, registers and quantum memories across a wide range of platforms, ranging from implementation of parametric driving in superconducting circuits~\cite{Royer2018,Eddins2019} and optomechanical devices~\cite{Aspelmeyer2014} to the coupling of microwave resonators to ensembles based on electron spins~\cite{Schuster2010},  nitrogen-vacancy centres~\cite{Kubo2010}, and semiconductor qubits~\cite{Mi2017}. 
{In recent studies, parametric modulation of potential has been used to observe enhanced effective interaction in trapped ions~\cite{Burd2021}, while superradiant phase transition and strong entanglement have been observed in a nuclear magnetic resonance quantum simulator using antisqueezing effects~\cite{Chen2021b}. }

Therefore, the protocol can be a powerful mechanism for designing versatile devices with significantly reduced error in information processing, and with much simpler engineering and low experimental overheads. Notably, the parametrically driven system can also be used to experimentally investigate regimes with ultrastrong coupling in ensembles, which will not only allow for the study of interesting protocols in quantum information but also throw more light on fundamental physics related to collective phenomena, many-body entanglement and nonequilibrium, driven-dissipative dynamics.\\

\noindent {\emph{Note.--} The datasets and source code employed in both the mean-field and variational renormalization group methods are available online~\cite{SimCodes}.}

\begin{acknowledgments}
We thank Sudipto Singha Roy and Rahul Gupta for helpful suggestions and acknowledge the use of SpaceTime, the high-performance computing facility at IIT Bombay, for a part of the quantum simulations. H.S. acknowledges financial support from the Prime Minister’s Research Fellowship (ID: 1302055), Govt. of India. H.S.D. acknowledges funding from SERB-DST, India under Core-Research Grant (CRG/2021/008918) and IRCC, IIT Bombay (RD/0521-IRCCSH0-001). 
\end{acknowledgments}

\appendix
\section{Simulation Parameters}\label{App1}
The parameters used in our mean-field and variational renormalization group methods are comparable to those observed across different hybrid systems.  
Table~\ref{Tab:Params} compares the cavity decay rates ($\kappa$), spin loss rate ($\gamma_{h}$), spin dephasing rate ($\gamma_{p}$), effective coupling strength ($\Omega$), and inhomogeneity ($\delta$) used in our numerical study alongside values reported in experiments and theory for a host of different platforms. The values chosen in our simulations are of similar orders of magnitude as those in these studies. We note that our framework is agnostic to any specific system, and as such, can be used to simulate the dynamics across all the platforms mentioned in the table.

\bibliographystyle{quantum}
\bibliography{references}

\begin{thebibliography}{10}
\bibitem{Ecomment}
The envelope is taken for representation as the cavity state regains all its coherence only at specific times. The actual fidelity oscillates as information is coherently exchanged between the spin ensemble and the cavity modes.
\end{thebibliography}
\end{document}